\documentstyle[aps,floats,psfig,twocolumn]{revtex}
\def\gz{\ifmmode{Z\hskip -4.8pt Z}
    \else{\hbox{$Z\hskip -4.8pt Z$}}\fi} 

\newcommand{\be}{\begin{equation}}
\newcommand{\ee}{\end{equation}}
\newcommand{\bea}{\begin{eqnarray}}
\newcommand{\eea}{\end{eqnarray}}

\newcommand{\rd}{\mbox{d}}

\newcommand{\re}{\mbox{e}}

\begin{document}
\tighten
\draft
\title{Phase diagram and optical conductivity of the ionic Hubbard model}

\author{Ph. Brune $^{a}$, G.~I. Japaridze $^{a,b}$, A.~P. Kampf $^{a}$, and M. Sekania $^{a}$}
\address{
$^a$ Institut f\"ur Physik, Theoretische Physik III,
Elektronische Korrelationen und Magnetismus,\\
Universit\"at Augsburg, 86135 Augsburg, Germany\\
$^b$ Institute of Physics, Georgian Academy of Sciences,
Guramishvili Str. 6, 380077, Tbilisi, Georgia}
\address{~
\parbox{14cm}{\rm 
\medskip
We investigate the ground-state phase diagram of the one-dimensional ``ionic'' 
Hubbard model with an alternating periodic potential at half-filling by the 
bosonization technique as well as by numerical diagonalization of finite
systems with the Lanczos and density matrix renormalization group (DMRG) 
methods. Our results support the existence of a single ``metallic'' transition 
point from a band to a correlated insulator with simultaneous charge and 
bond-charge order. In addition, we present results for the optical conductivity
obtained by the dynamical DMRG method. The insulator-insulator phase transition
scenario is discussed in detail including a critical review of existing 
approaches and results for the ionic Hubbard model.
}}

\maketitle

\section{Introduction}
For more than two decades the correlation induced metal-insulator transition 
(MIT) and its characteristics has been one of the challenging problems 
in condensed matter physics \cite{Imada}. This MIT is often accompanied by a
symmetry breaking and the development of long range order \cite{Lee}. In 
dimension D=1 this ordering can only be connected with the breaking of a 
discrete symmetry. Examples include commensurate charge density waves (CDWs)
and Peierls dimerization phenomena. The MIT can be driven either by varying the
electron density or the electron-electron interaction strength. If the MIT is 
approached from the metallic side, the electrical conductivity or the
electronic compressibility may be used to characterize the transition; the 
approach of the transition from the insulating side may on the other hand be 
properly captured by the divergent behavior of the electric susceptibility 
\cite{Aebischer}. 

In addition, the features of the insulating phase may change qualitatively with
the interaction strength, allowing for quantum phase transitions 
between insulating phases. The extended 
Hubbard model at half-filling with an on-site ($U$) and a nearest neighbor 
($V$) Coulomb repulsion provides a prominent example with a transition from a 
Mott insulator (MI) to a CDW insulator in the vicinity of the 
$U=2V$ line in the phase diagram \cite{Solyom}. Remarkably, the transition
involves an intermediate phase with a bond-order wave (BOW) 
\cite{Nakamura,Campbell}. 

In recent years particular attention has been given to another example for an 
extension of the Hubbard model which includes a staggered potential term 
\cite{Nagaosa86,Fabrizio99,Gidopoulos99}. The corresponding Hamiltonian has 
been named the ``ionic Hubbard model'' (IHM); at half-filling this model 
undergoes a different kind of insulator-insulator transition from a band to a 
correlated Mott-Hubbard type insulator. Conflicting results have so far been 
reported regarding the nature of the transition, the possibility of two rather
than one critical point, or the appearance of BOW order
\cite{Resta95,Fabrizio99,Gidopoulos99,Wilkens00,Pati00,Torio01}. 
Given the numerous unresolved issues we reinvestigate in detail 
the IHM using both, numerical and analytical tools.
Specifically we study the 1D Hamiltonian 
\begin{eqnarray}
 \nonumber
 H&=&-t\sum_{i,\sigma}(1+(-1)^i\delta)\left(
             c^{\dagger}_{i\sigma}c^{\phantom{\dagger}}_{i+1\sigma}
            +H.c.\right)\\
   &&+U\sum_i n^{\phantom{\dagger}}_{i\uparrow}
              n^{\phantom{\dagger}}_{i\downarrow}
     +{\Delta\over 2}\sum_{i\sigma} (-1)^i
            n^{\phantom{\dagger}}_{i\sigma}
 \quad,
 \label{IonicHam}
\end{eqnarray}
where $c^{\dagger}_{i\sigma}$ creates an electron on site $i$ with spin
$\sigma$, $n^{\phantom{\dagger}}_{i\sigma}=c^{\dagger}_{i\sigma}
c^{\phantom{\dagger}}_{i\sigma}$. 
$\Delta$ is the potential energy difference between neighboring sites, and
$\delta$ a Peierls modulation of the hopping amplitude $t$. In the limit
$\Delta=\delta=0$, Eq. (\ref{IonicHam}) reduces to the ordinary
Hubbard model, the limit $\Delta=0$ and $\delta>0$ is called the
Peierls-Hubbard model (PHM), and the limit $\Delta>0$ and $\delta=0$ is usually
referred to as the IHM. In the following, we will focus mainly on the
effect of the on-site modulation $\Delta$, so we implicitly assume $\delta=0$
except where stated otherwise.

The IHM was first proposed and discussed almost 20 years ago in the context of
organic mixed-stack charge-transfer crystals with alternating donor ($D$) and
acceptor ($A$) molecules ($\dots D^{+\rho}A^{-\rho}\dots$)
\cite{Torrance81,Nagaosa86}. These stacks form quasi-1D insulating chains, and
are classified into two categories depending on the amount of charge transfer
$\rho $: quasi-neutral for $\rho<0.5$, and quasi-ionic for $\rho>0.5$.
Torrance {\it et al.} \cite{Torrance81} found that at room temperature and
ambient pressure, these materials are either mostly ionic ($\rho\approx 1$) or
mostly neutral ($\rho\approx 0$), but several systems undergo a reversible
neutral to ionic phase transition (NIT) \cite{Mitani84}, i.e. a discontinuous
jump in the ionicity $\rho$ upon changing temperature or pressure with the
simultaneous appearance of a Peierls lattice distortion
\cite{Girlando83,LeCointe95}. The original modelling of the NIT in
charge-transfer salts included different Coulomb repulsion parameters on A and 
D molecules as well as intersite interactions \cite{Horovitz,Girlando86} and 
electron-molecular vibration coupling \cite{Painelli88}. In these extended 
ionic Hamiltonians the NIT was found to be first or second order depending on 
the relative strengths of the different interactions. 

In addition, the IHM has been used in a quite different context to describe the
ferroelectric transition in perovskite materials such as BaTiO$_3$
\cite{Egami93,Ishihara94,Resta95} or KNbO$_3$ \cite{Neumann92}. The IHM here is
used as a simple model for a binary compound, $AB$, with the $A$ sites 
representing the transition metal element (e.g. Ti), and the $B$ sites
representing the oxygen. The staggered potential in this case is the energy 
difference of the levels on the $A$ and $B$ sites \cite{Egami93,Ishihara94}. 

The results of our work on the IHM support the existence of a single transition
at a critical coupling $U_c(\Delta)$ from a band- to a correlated insulator 
(CI) phase. The transition  originates from a ground-state level crossing with 
a change of the site-parity eigenvalue. From our results we identify the
strong coupling phase at $U>U_c(\Delta)$ as a CI with long-range CDW order. 
Two possibilities arise for the unconventional CI phase: it either has a
vanishing spin excitation gap and exhibits on top of a CDW pattern an 
identical power law decay of SDW and dimer-dimer correlations; or the spin gap 
remains finite in the CI phase with true long range BOW order coexisting with 
a CDW. Within the achievable numerical accuracy our data support the latter 
scenario.Furthermore, we provide the first numerical evaluation of the 
optical conductivity for the IHM which allows for an additional 
characterization of the ''metallic'' transition point $U_c(\Delta)$ where the 
optical gap closes. The distinction between charge and optical gaps is crucial 
for the structure of the ground-state phase diagram.

The outline of our paper is as follows: In sections II and III we summarize 
previously proposed scenarios for the ground-state phase diagram using 
bosonization and symmetry arguments. Section IV contains a detailed reanalysis 
of the bosonization approach to the IHM. In sections V and VI we present our 
numerical Lanczos and DMRG results, from which we draw conclusions for the 
ground state phase diagram, and in section VII we discuss dynamical DMRG 
results for the optical conductivity. Finally, we conclude and summarize in 
section VIII.

\section{Insulator-insulator transition}

A good starting point for understanding the existence of a phase transition in
the IHM is the atomic limit \cite{McConnell,Gidopoulos99}. For
$t=0$, it is immediately seen that at half-filling and $U<\Delta$ the
ground-state of the IHM has two electrons on the odd sites, and no electrons
on the even sites corresponding to CDW ordering with maximum amplitude. On
the other hand, for $U>\Delta$ each site is occupied by one electron.
Obviously, for $t=0$ a transition occurs at a critical value $U_c=\Delta$.
This transition is expected to persist for finite hopping amplitudes
$t>0$, where the alternating potential still defines two sublattices, doubling 
the unit cell and opening up a band gap $\Delta$ for $U=0$ at
$k=\pm\pi/2$. For $t>0$ the critical coupling shifts to $U_c(t)>\Delta$, where
$U_c$ increases monotonically with increasing $\Delta$. For $U,\Delta\gg t$ 
the system is close to the atomic limit, and $U_c$ approaches $\Delta$ from
above.

For $U=0$ the ground state at half-filling is a CDW band insulator (BI), whose 
elementary spectrum consists of particle-hole excitations over the band gap. We
consider a system to be in a BI phase when the criterion $\Delta_S=\Delta_C$ 
holds, where the spin ($\Delta_S$) and the charge gap ($\Delta_C$) are given by
\begin{eqnarray}
 \nonumber
 \Delta_S &=& E_0(N=L,S_z=1)-E_0(N=L,S_z=0)\, ,\\ \nonumber
 \Delta_C &=& E_0(N=L+1,S_z=1/2)\\\nonumber
          &+& E_0(N=L-1,S_z=1/2)-2E_0(N=L,S_z=0).
 \label{Gaps1}
\end{eqnarray} 
$E_0(N,S_z)$ is the ground-state energy, $L$ the system length, $N$ the number 
of electrons, and $S_z$ the $z$-component of the total spin. For $U>U_c$ and
$U\gg t,\Delta$, the charge gap is set by the Coulomb interaction 
$U$, and the system is a CI with $\Delta_C>\Delta_S$. However, in contrast to 
the cases with $\Delta=0$ or $t=0$, CDW order is expected for all
finite values of $U$.

The IHM is distinctly different from the PHM. This is also a 
BI at $U=0$, but in contrast to the IHM has $\Delta_C>\Delta_S>0$ for any 
value of $U>0$, i.e. the phase transition from the Peierls BI to the CI occurs 
at $U_c=0$.

A renewal of interest in the IHM started with the bosonization analysis 
of Fabrizio {\it et al.} (FGN) \cite{Fabrizio99} and exact diagonalization 
studies \cite{Gidopoulos99}, where a new scenario for the intermediate region 
$U\approx\Delta$ was proposed. Several subsequent numerical treatments of the 
model \cite{Takada00,Qin00,Wilkens00,Pati00,Torio01} led to partially 
contradicting results. Different conclusions were reached about the nature of
the insulator-insulator phase transition, the possible existence of two 
transitions, and the question whether the spin gap closes in the strong 
coupling phase.

FGN argued that for small but finite $U$ the BI persists up to a critical value
$U_{c1}$, where $\Delta_C=0$ and the system might be ``metallic''. Upon further
increasing $U$ they predicted a ``spontaneously dimerized'', (SDI or 
equivalently a BOW) intermediate phase, which should undergo a continuous 
transition into the MI phase at a second critical value $U_{c2}>U_{c1}$. A BOW 
ground state is characterized by long-range staggered bond-density correlations
\begin{eqnarray}
 g_{B}(r)&=&{1\over L}\sum_{i}\,\langle\psi_0|\,B_iB_{i+r}\,|\psi_0\rangle
 \quad, \\
 B_i&=&\sum_\sigma\left( c^{\dagger}_{i\sigma}c^{\phantom{\dagger}}_{i+1\sigma}
                      +H.c.\right)
\end{eqnarray}
and implies a finite spin gap $\Delta_S>0$. 

Evidence for such a BOW state has been reported by Wilkens and Martin 
in a quantum Monte Carlo (QMC) study \cite{Wilkens00}. More 
precisely, the authors of Ref. \cite{Wilkens00} observed a single transition 
from the BI to the correlated BOW phase. On the contrary, DMRG results 
\cite{Qin00} and Lanczos studies of level crossings in the excitation spectra 
support the existence of two transitions with an intervening BOW state 
\cite{Torio01}. From finite size extrapolations it was furthermore concluded 
that $\Delta_S=0$ above the critical region \cite{Resta95,Ortiz96,Qin00}.
Whether CDW order persists also in the strong coupling 
regime remained unsettled in previous work.

\section{Symmetry analysis}

Insight into the nature of the BI-MI transition is obtained from a symmetry
analysis. The IHM is invariant with respect to {\it inversion} at a site and
{\it translation} by two lattice sites. Thus, any nondegenerate eigenstate of
$H$ is also an eigenstate of the operators that generate the corresponding
symmetry transformation. If we denote the site inversion operator by $P$,
defined through
\begin{equation}
 Pc^{\dagger}_{i\sigma}P^{\dagger}=c^{\dagger}_{L-i\sigma}
 \quad\mathrm{for} \;i=0, \cdots, L-1
 \quad,
\end{equation}
and ${\hat T}_j$ for a translation
by $j$ sites, then any nondegenerate eigenstate $|\psi_n\rangle$ of $H$ must
obey $P|\psi_n\rangle=\pm|\psi_n\rangle$ and
${\hat T}_2|\psi_n\rangle=|\psi_n\rangle$. Because $[H,{\hat T}_1]\neq 0$, a
non-degenerate eigenstate $|\psi_n\rangle$ of $H$ can not be an eigenstate
of ${\hat T}_1$.

For the half-filled Hubbard model ($\Delta=\delta=0$) the site-parity
eigenvalue of the ground state can be determined in the limits $U=0$
and $U\gg t$ \cite{Gidopoulos99}. For $U=0$, the ground state is a direct 
product of spin up and spin down Slater determinants, both formed from the 
same occupied spatial wavefunctions with the same parity $P_\sigma=\pm 1$, 
so the parity eigenvalue of the total wavefunction is given by their product
$P=P_\downarrow\times P_\uparrow =\pm 1$.
Hence, the state at $U=0$ is even under site inversion. On the other hand, in 
the large $U$ limit the mapping to the Heisenberg Hamiltonian can be used to 
show that for $L=4n$ with periodic boundary conditions (PBC) or $L=4n+2$ with 
antiperiodic boundary conditions (APBC) the ground state obeys 
$P|\psi_0\rangle=-|\psi_0\rangle$ (for details see \cite{Gidopoulos99}). 
In finite chains, these combinations of chain lengths and boundary conditions
ensure that $k=\pi/2$ is an allowed momentum in the Brillouin zone (BZ).
This is important at half-filling, where $k=\pm\pi/2$ is the BZ boundary. It 
follows that upon increasing the number of sites $L$, the ground state for 
$U\gg t$ will be odd with respect to $P$ as long as $k=\pi/2$ is an allowed 
$k$ value, and this feature will persist in the thermodynamic limit 
$L\rightarrow+\infty$. For the ordinary Hubbard model the discrete inversion 
symmetry changes therefore between $U=0$ and $U\gg t$. Since the
model has no phase transition for $U>0$, the ground state has $P=+1$ 
only for $U=0$, and $P=-1$ for any $U>0$.

However, in the IHM the phase transition from a BI to a CI occurs at some 
finite $U_c>0$. This suggests 
that the parity of the ground state remains even not only for $U=0$, but for 
all $U<U_c$. At $U_c$, a ground-state level crossing occurs on finite chains, 
as confirmed by exact diagonalization studies (see below), connected with a 
site-parity change. 

In the strong-coupling limit, the IHM was argued in some previous treatments 
\cite{Gidopoulos99,Fabrizio99,Torio01} to be a MI with $\Delta_S=0$. For the 
IHM with $U\gg t,\Delta$ the effective Heisenberg spin model 
\begin{equation}
 H_{eff}=J\sum_{i}{\bf S}_i\cdot{\bf S}_{i+1}
         +J'\sum_{i}{\bf S}_i\cdot{\bf S}_{i+2}
\label{SpinHam}
\end{equation}
was derived to describe the low-energy physics \cite{Nagaosa86}. In Eq. 
(\ref{SpinHam}) the exchange couplings are given by
\begin{eqnarray}
 \nonumber
 J      &=& \frac{4t^2}{U}\left[\frac{1}{1-x^2}
            -\frac{4t^2}{U^2}\frac{1+4x^2-x^4)}{(1-x^2)^3}\right]\,\, ,\\
 J'     &=& \frac{4t^4}{U^3}\frac{(1+4x^2-x^4)}{(1-x^2)^3}
 \label{JCouplings}
 \quad,
\end{eqnarray}
where $x=\Delta/U$. It is important to recall that in the derivation of the
effective spin model the Hilbert space of the IHM is divided into two 
subspaces -- one containing no doubly occupied sites, and one containing all 
states with at least one doubly occupied site. The atomic limit of the 
IHM is taken as the unperturbed problem $H_0$, and the kinetic term in 
(\ref{IonicHam}) is treated as a pertubation. The ground state of $H_0$ is 
therefore in the subspace without double occupancies. The kinetic term lifts 
the spin degeneracy in this subspace. Introducing the projection operator
$P=\prod_i \left(1-n^{\phantom{\dagger}}_{i\uparrow}
n^{\phantom{\dagger}}_{i\downarrow}\right)$, the effective Hamiltonian 
$H_{eff}=P\exp(iS)H\exp(-iS)P$ is obtained by a canonical transformation $S$ 
and an expansion in powers of $t/U$ around $H_0$ 
\cite{MacDonald88,Takahashi77}.

The result (\ref{SpinHam}) would imply that in the strong-coupling limit of the
IHM the low-energy physics is qualitatively similar to that of the Hubbard 
model, with modified exchange coupling constants $J$ and $J'$. 
For next-nearest neighbor couplings $J'<0.24J$ the spin gap vanishes 
\cite{Haldane}; thus the system would be a true MI. The coupling constants
(\ref{JCouplings}) satisfy the condition $J'<0.24J$ at least for $U>3.6t$ for
$\Delta\le t$ and $U>3.6\Delta$ for $\Delta>t$. However, the effective model 
(\ref{SpinHam}) is invariant with respect to translations by {\it one} lattice 
spacing, whereas the original IHM is invariant only with respect to
translations by {\it two} lattice spacings. Thus, the effective spin
Hamiltonian has a higher symmetry than the original model from which it was 
derived, and the arguments of the strong-coupling expansion in favor of 
$\Delta_S=0$ for $U>U_c$ cannot be considered rigorous as was also pointed out 
in Ref. \cite{Wilkens00}.

Of course, also in the Hubbard chain for any $U<\infty$ the double
occupancy is not zero. However, these ``virtual doublons'' are distributed 
equally over all sites, and therefore do not change the translational symmetry.
If only the spin excitations are considered, these doublons are never traced.

As originally discussed in Ref. \cite{Girlando86} the existence of a level
crossing on finite chains at a critical $U_c$ leaves the following 
possibilities for the ground state of the IHM at $U>U_c$ in the thermodynamic 
limit: 1. The ground state remains unique with site-parity quantum number 
$P=-1$. This necessarily implies the absence of BOW and the system is in the 
MI phase with a vanishing spin gap and absence of long range CDW. Remarkably 
and left uncommented before, this scenario implies the highly unusual 
situation that the strong coupling ground state has a higher symmetry than the 
IHM Hamiltonian. 2. The ground state is twofold degenerate and supports a 
long-range BOW+CDW order. This possibility requires the spin gap to remain
finite \cite{Fabrizio99}. These intimately connected properties are violated 
in the conclusions reached in Ref. \cite{Pati00}. This second 
scenario though still allows for an additional second transition into the MI 
phase as characterized in 1.

\section{Bosonization results}

The standard bosonization procedure allows to express the Hamiltonian density 
in the following way \cite{Fabrizio99}: ${\cal H}={\cal H}_{c}+{\cal H}_{s}+
{\cal H}_{cs}$, where the charge and spin degrees of freedom are represented by
a sine-Gordon model for scalar fields  $\phi_{c}$ and $\phi_{s}$, respectively:
\begin{eqnarray}
{\cal H}_{c}&=&{v_{c}\over 2}[P^2_{c}(x)+(\partial_x\phi_{c})^2]\nonumber\\
&-&\frac{U}{2\pi a_{0}^{2}} \cos\left(\sqrt{8\pi K_{c}}\phi_{c}(x)\right) ,
\label{SinGorC}\\  
{\cal H}_{s}&=&{v_{s}\over 2}[P^2_{s}(x)+(\partial_x\phi_{s})^2]\nonumber\\ 
\label{SinGorS}
&+&\frac{U}{2\pi a_{0}^{2}} \cos\left(\sqrt{8\pi}\phi_{s}(x)\right) \, ;
\end{eqnarray}
$a_0$ is the lattice constant. The ionic term determines the spin-charge 
coupling
\begin{equation}
{\cal H}_{cs}=\frac{\Delta}{\pi a_{0}}
\sin\left(\sqrt{2\pi K_{c}}\phi_{c}(x)\right)\cos\left(\sqrt{2\pi}\phi_{s}(x)
\right) . 
\label{SGcs}
\end{equation}
Here $P_{c (s)}(x)$ is the momentum conjugate to $\phi_{c (s)}(x)$, $v_{c (s)}$
is the velocity of charge (spin) excitations and $K_{c} \simeq 1-U/4 \pi t $.  
This means that the model (\ref{SGc}) is in a strong-coupling regime for 
arbitrary $U>0$, and at $\Delta =0$ the dynamically generated mass determines 
the charge gap in the system.

\subsection{Forward scattering model with ionic and/or Peierls distortion}

We first consider the limit of a weak Hubbard interaction  
$U \ll \Delta \ll t$, where the properties of the system are mostly determined 
by the ionic distortion. In the infrared (low energy - large distance) 
limit the Umklapp and backward scattering processes, described in Eqs.
(\ref{SinGorC})-(\ref{SinGorS}) by the terms 
$U\cos\left(\sqrt{8\pi K_{c}}\phi_{c}(x)\right)$ and 
$U\cos\left(\sqrt{8\pi}\phi_{s}(x)\right)$, respectively, are frozen out.
Therefore, we initially neglect these terms and consider the Hamiltonian 
\begin{eqnarray}
{\cal H}_{TL}&=&\int{\rm d}x\Big\{{v_{c}\over 2}[P^2_{c}(x)+(\partial_x
\phi_{c})^2]\Big\} \\\nonumber
&&+\int{\rm d}x\Big\{ {v_{s}\over 2}[P^2_{s}(x)+(\partial_x
\phi_{s})^2]\Big\}\quad\\\nonumber
&-&\frac{\Delta}{\pi a_{0}}\int{\rm d}x
\sin\left(\sqrt{2\pi K_{c}}\phi_{c}\right)
\cos\left(\sqrt{2\pi K_{s}}\phi_{s}\right)\quad\, ,
\label{ionTL}
\end{eqnarray}
i.e. the half-filled Tomonaga-Luttinger (TL) model with ionic distortion. The 
non-interacting system corresponds to the particular limit $K_{c}=K_{s}=1$. 
For generality we do not assume $SU(2)$ symmetry of the spin channel and 
therefore the parameter $K_{s}$ is not fixed to unity. 

If we decouple the interaction term in a mean-field manner by introducing 
\begin{eqnarray}
\label{m_{c}}
m_{c}&=& \Delta \cdot \langle \cos(\sqrt{2\pi K_{s}}\phi_{s})\rangle\,\, ,\\
\label{m_{s}}
m_{s}&=& \Delta \cdot \langle \sin(\sqrt{2\pi K_{c}}\phi_{c})\rangle\, ,
\end{eqnarray}
the bosonized Hamiltonian reads ${\cal H}={\cal H}_c+{\cal H}_s$ where
\begin{eqnarray}
{\cal H}_{c}&=&\int dx\Big\{{v_{c}\over 2}[P^2_{c}(x)+(\partial_x
\phi_{c})^2]\nonumber\\
\label{FreeBosMF}
&&-\frac{m_{c}}{\pi a_{0}}\sin(\sqrt{2\pi K_{c}} \phi_{c})\Big\},\label{SGc}\\ 
{\cal H}_{s}&=&\int dx\Big\{{v_{s}\over 2}[P^2_{s}(x)+(\partial_x
\phi_{s})^2]\nonumber\\
&&-\frac{m_{s}}{\pi a_{0}}\cos(\sqrt{2\pi K_{s}} \phi_{s})\Big\}.
\label{SGs}         
\end{eqnarray}
We can estimate the renormalized masses as 
\bea
M_{c} & \simeq &\Lambda \left({m_{c}\over\Lambda}\right)^{2/(4-K_c)}
\,\, ,\\
M_{s} & \simeq & \Lambda\left({m_{s}\over\Lambda}\right)^{2/(2-K_s)},
\eea
where $\Lambda$ is a cut-off energy. Similarly, the expectation values 
of $\sin(\sqrt{2\pi K_{c}}\phi_{c})$ and 
$\cos(\sqrt{2\pi K_{s}} \phi_{s} )$) are estimated as
\bea
\langle \sin(\sqrt{2\pi K_{c}}\phi_{c})\rangle &\simeq & 
\left({m_{c}\over\Lambda}\right)^{K_c/(4-K_c)}\, ,\\
\langle \cos(\sqrt{2\pi K_{s}} \phi_{c})\rangle &\simeq & 
\left({m_{s}\over\Lambda}\right)^{K_s/(4-K_s)}.
\eea
The self-consistency equation 
\be\label{SConsistEq}
\Delta=\frac{m_{c} }{\langle \sin(\sqrt{2\pi K_{s}}\phi_{s})\rangle}= \frac{m_{s}}{\langle \sin(\sqrt{2\pi K_{c}}\phi_{c})\rangle } 
\ee
leads to
\be\label{SConsistEqR1}
{m_s\over\Lambda}= 
\left({m_c\over\Lambda}\right)^{(4-K_s)/(4-K_c)}.
\ee
Using Eqs. (\ref{m_{c}})-(\ref{SConsistEq}) one easily finds that 
$$
M_{s}=M_{c}=\Lambda \left({\Delta\over\Lambda}\right)^{2/(4-K_c-K_s)}\, ,
$$
and the system is a BI with a weakly renormalized value 
of the excitation gap. In particular, for the non-interacting system 
$K_{c}=K_{s}=1$ and $M_{s}=M_{c}=\Delta$. 

For the PHM the bosonized expression for the Peierls distortion term is 
given by 
\be
-\frac{\delta}{\pi a_{0}}
\cos\left(\sqrt{2\pi}\phi_{c}\right)\cos\left(\sqrt{2\pi}\phi_{s}\right).
\ee
By following again the same steps as above the half-filled TL
model with a Peierls distortion also exhibits the 
properties of a BI with equal charge and spin gaps.
 
\subsection{Renormalization effects from the short-range interaction}

In order to account for the renormalization of the BI gap by the 
short-range part of the interaction, we consider again the limiting case 
$U \ll \Delta \ll t$. We treat both the ionic term and the Hubbard 
interaction on the same footing in a perturbative way. For this purpose it is 
convenient to consider the 2D Euclidean action:
\bea
&S&= S_{0} + S_{{\it int}}\,\, , \\
&S_{0}&=v_{F} \int \rd^2 {\bf x}~\Big\{\frac{1}{2}[ (\nabla \phi_{c})^2 + 
(\nabla \phi_{s})^2]\Big\}\,\, ,
\nonumber\\
&S_{{\it int}}&=v_{F} \int \rd^2 {\bf x}~\Big\{\frac{m_{cs}}{\pi a_{0}}
\sin \left(\sqrt{2\pi K_{c}} \phi_{c}\right)\cos\left( 
\sqrt{2\pi} \phi_{s}\right)\nonumber\\
&&+ \frac{M_c}{\pi a_{0}^{2}} 
\cos \left(\sqrt{8\pi K_{c}} \phi_{c}\right) + 
\frac{M_s}{\pi a_{0}^{2}}
\cos \left( \sqrt{8\pi} \phi_{s}\right)
\Big\}\, .\nonumber
\eea
Here the dimensionless coupling constants are given by
\be
M_{c}=-\frac{U}{4\pi t},\qquad M_{s}=\frac{U}{4\pi t},\qquad m_{cs}
=-\frac{\Delta}{2\pi t}.
\\\label{McKc}  
\ee
Expanding the partition function  
\begin{equation}
Z=\int \mbox{D} \Phi_{s}\mbox{D}\Phi_{c} \re^{- S_{0} [\Phi_{c},\Phi_{s}]}
[1 - S_{{\it int}} +\frac{1}{2}S^{2}_{{\it int}} + ... ]
\label{Expansion}
\end{equation}
we integrate out all configurations in the second order term with 
$| {\bf x}- {\bf x'}| \sim a_{0}$. Using the operator product 
expansion formulas
\begin{eqnarray}
&&\int \frac{\rd^2 {\bf x}}{a_{0}^{2}}\int 
\frac{\rd^2 {\bf x'}}{a_{0}^{2}}
\sin \left(\phi({\bf x})\right)\cdot\cos\left(2\phi({\bf x'})\right) =
\nonumber\\
&-&\frac{\pi}{2}\int \frac{\rd^2 {\bf x}}{a_{0}^{2}}[
\sin\left(\phi_{c}({\bf x})\right) + ... ]\,\, ,  
\label{OPE1}\\
&&\int \frac{\rd^2 {\bf x}}{a_{0}^{2}}\int 
\frac{\rd^2 {\bf x'}}{a_{0}^{2}}
\cos \left(\phi({\bf x})\right)\cdot\cos\left(2\phi({\bf x'})\right) =
\nonumber\\
&+&\frac{\pi}{2}\int \frac{\rd^2 {\bf x}}{a_{0}^{2}}[
\cos\left(\phi_{c}({\bf x})\right) + ... ]\,\, ,\label{OPE2}\\
&&\int \frac{\rd^2 {\bf x}}{a_{0}^{2}}
\int \frac{\rd^2 {\bf x'}}{a_{0}^{2}}
\cos \left(\phi({\bf x})\right)\cdot\cos\left(\phi({\bf x'})\right) =
\nonumber\\
&+&\frac{\pi}{2}\int\frac{\rd^2 {\bf x} }{a_{0}^{2}}[1 - \frac{1}{2} 
(\nabla \phi_{c})^2 + \cos\left(2\phi({\bf x})\right)+...]\,\, ,\label{OPE3}\\
&&\int \frac{\rd^2 {\bf x}}{a_{0}^{2}}
\int \frac{\rd^2 {\bf x'}}{a_{0}^{2}}
\sin \left(\phi({\bf x})\right)\cdot\sin\left(\phi({\bf x'})\right) =
\nonumber\\
&+&\frac{\pi}{2}\int\frac{\rd^2 {\bf x} }{a_{0}^{2}}[1 - \frac{1}{2} 
(\nabla \phi_{c})^2 - \cos\left(2\phi({\bf x})\right)+...]\,\, ,
\end{eqnarray}
where dots denote strongly irrelevant terms of higher critical 
dimensionality, we obtain the following renormalization of the 
model parameters:
\begin{eqnarray}
\tilde{\Delta}& = &\Delta\left(1-\lambda {U\over t}\right), \\\label{IRgaps1}
\tilde{M_{c}} &=& -U\left(1 - \lambda_c{\Delta^2\over Ut}\right), \\\label{IRgaps2}
\tilde{M_{s}} &=& U\left(1- \lambda_s{\Delta^2\over Ut}\right).\label{IRgaps3}
\end{eqnarray}
$\lambda,\lambda_c$, and $\lambda_s$ are positive numbers of order unity. 
We observe that

${\bullet}$ the parameter of the ionic distortion is renormalized linearly by 
the on-site repulsion;

${\bullet}$ the amplitudes of the Umklapp  and backward scattering processes 
($\tilde{M}_c$ and $\tilde{M}_s$, respectively) are renormalized quadratically 
by the ionic distortion;

${\bullet}$ the $U \leftrightarrow -U$ asymmetry of the model is clearly 
manifested: at $U > 0$ the amplitude of the strongly relevant Umklapp 
scattering decreases, while at $U<0$ the amplitude of the strongly relevant 
backscattering processes increases.

Therefore we conclude that in the IHM the BI phase is more stable for a 
repulsive than an attractive Hubbard interaction.

For a Peierls distortion the effect of the electron-electron interaction at 
short distances is rather different. Assuming $U \ll\delta\ll t$ the same 
procedure as above gives in this case the following expressions for the 
renormalized parameters
\begin{eqnarray}\label{PRgaps1P}
\tilde{\delta}& = &\delta\left(1 + O\left({U\over t}\right)^2\right),\\
\label{IRgaps2P}
\tilde{M^{P}_{c}} &=& -U\left(1 +\lambda^{\prime}_{c}{\delta^2\over Ut}\right), \\
\label{PRgaps1}
\tilde{M_{s}^{P}} &=& U\left(1- \lambda^{\prime}_{s}{\delta^2\over Ut}\right).\label{IRgaps3P} 
\end{eqnarray}
Therefore, contrary to the ionic case,

${\bullet}$ the amplitude of the Peierls distortion is not renormalized to 
linear order in $U$;

${\bullet}$ the amplitudes of the Umklapp and backward scattering processes 
are renormalized quadratically by the Peierls distortion term;

${\bullet}$ at $U>0$ the amplitude of the strongly relevant Umklapp scattering 
increases, while at $U<0$ the amplitude of the strongly relevant 
backscattering processes increases.

Based on these arguments, one is led to predict that for the PHM
the deviation from the BI behavior due to the on-site coupling of 
arbitrary sign could be similar to that of the attractive IHM.  

Therefore we conclude that the BI phase in the Peierls-Hubbard model is less 
stable against a Peierls distortion which 
is not renormalized to linear order in $U$, and the amplitudes of the relevant 
scattering processes $M^{P}_{c}$ at $U>0$ and $M^{P}_{s}$ at $U<0$ always 
increase. The deviation 
from the BI behavior in the case of the attractive IHM and the 
PHM should therefore happen almost at the same values of the on-site 
interaction. The fundamentally different behavior of the repulsive IHM and 
the PHM will indeed be verified below.

\subsection{The FGN phase diagram}

In Ref. \cite{Fabrizio99} FGN proposed a new scenario for the 
BI to MI crossover in the IHM. The key ingredient of their theory is the 
presence of {\em two separate transitions}: an Ising-type transition at 
$U^{c}_{ch}$ where the charge gap vanishes and a continuous transition at 
$U^{c}_{sp}>U^{c}_{ch}$ where the spin gap vanishes. The charge excitations 
are gapped for arbitrary $U\neq U^{c}_{ch}$, while the spin sector remains 
gapless at $U>U^{c}_{sp}$. The ground-state phase diagram of the IHM and the 
properties of the charge and spin gapped phases were argued to be 
qualitatively captured by the effective potential: 
\bea
{\cal V} \left( \phi_c ,\phi_s  \right)=&-& \tilde{\Delta} \sin 
\left(\sqrt{2\pi K_{c}}\phi_c \right)\cos \left(\sqrt{2\pi}\phi_s \right)\\\nonumber 
&-&\tilde{M}_c \cos \left(\sqrt{8\pi K_{c}}\phi_c \right)  - 
\tilde{M}_s \cos \left(\sqrt{8\pi}\phi_s \right),\label{pot:general}
\eea
where $\tilde{M}_{c},\tilde{M}_{s},\tilde{\Delta}> 0$ are phenomenological 
parameters obtained by integrating out high-energy excitations. 
Minimizing ${\cal V} \left( \phi_c ,\phi_s  \right)$ with respect to
$\phi_c$ and $\phi_s$ results in the following sets of vacua: 
For $\tilde{\Delta}> 4 \tilde{M}_c$ (i.e. $U<U^{c}_{ch}$) 
\begin{eqnarray}
\label{vacDelta1a}
{\rm I.}\hskip0.1cm \sqrt{2\pi}\phi_s&=& 2 \pi n\, ,\,
\sqrt{2\pi K_{c}}\phi_c =  \frac{\pi}{2} 
 ~~ ({\rm mod~} 2\pi )\nonumber\\
{\rm II.}\hskip0.1cm \sqrt{2\pi}\phi_s &=& \pi +  2 \pi n,\, \,
\sqrt{2\pi K_{c}}\phi_c =  -\frac{\pi}{2} 
 ~~ ({\rm mod~} 2\pi )
\end{eqnarray}
while for $\tilde{\Delta}< 4 \tilde{M}_{c}$ (i.e. $U^{c}_{ch}<U<U^{c}_{sp}$)   
\begin{eqnarray}
{\rm I.}\hskip0.1cm \sqrt{2\pi}\phi_s&=& 2 \pi n\, ,\,
\sqrt{2\pi K_{c}}\phi_c = \varphi_0,~ \pi -
\varphi_0 ~~ ({\rm mod~} 2\pi )\nonumber\\
{\rm II.}\hskip0.1cm \sqrt{2\pi}\phi_s &=& \pi +  2 \pi n\, ,\,\nonumber\\
\sqrt{2\pi K_{c}}&\phi_c&= - \varphi_0,~ - \pi + \varphi_0 ~~
({\rm mod~} 2\pi) 
\label{g_c2}
\end{eqnarray} 
where $\varphi_{0}=\arcsin({\tilde{\Delta}}/4\tilde{M}_c)$. 

\begin{figure}[t!]
\centerline{\psfig{file=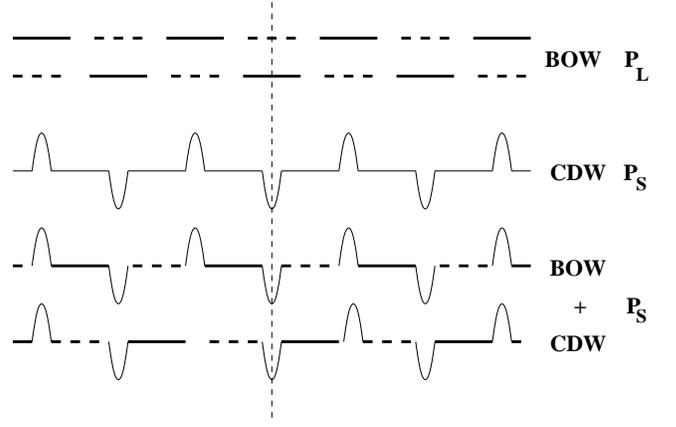,width=85mm,silent=}}
\vspace{2mm}
\caption{Illustration of the CDW + BOW phase. Top figure: two degenerate 
dimerization patterns; the vertical dashed line is a mirror axis indicating 
the link-reflection symmetry of each dimer state. Middle figure: schematic 
representation for 
a staggered CDW which has site-reflection symmetry. Lower
figure: the superposition of the CDW state with both dimerized patterns again 
allows for a site-reflection symmetric state.}
\label{fig:SDI}
\end{figure}

The symmetry properties of the various ordered phases are described by
the order parameters for  the short wavelength fluctuations of the 

${\bullet}$ {\it site}-located charge density wave:
\begin{eqnarray}\label{CDWop}
\Delta_{\small CDW} & = & (-1)^{n} \sum_{\sigma} c^{\dagger}_{n, \sigma}c_{n, \sigma}\nonumber\\
& \sim & \sin(\sqrt{2\pi K_{c}}\phi_{c}) \cos( \sqrt{2\pi}\phi_{s}) \,
\end{eqnarray}
${\bullet}$ {\it site}-located spin density:
\begin{eqnarray}
\Delta_{\small SDW} & = & {(-1)}^n \sum_{\sigma}\sigma \rho_{n, \sigma}\nonumber\\
& \sim & \cos(\sqrt{2\pi K_{c}}\phi_{c})
\sin(\sqrt{2\pi}\phi_{s}) \, ,\label{SDWop}   
\end{eqnarray}
${\bullet}$ {\it bond}-located charge density wave:
\begin{eqnarray}
\Delta_{\small BOW} &  = &  (-1)^{n}
 \sum_{\sigma}(c^{\dagger}_{n, \sigma}c_{n+1,\sigma} + h.c.)\nonumber\\ 
& \sim  & \cos(\sqrt{2\pi K_{c}}\phi_{c})  
\cos(\sqrt{2\pi}\phi_{s})\label{b-CDW}\, . 
\end{eqnarray}
For $U<U^{c}_{ch}$ the charge and spin excitation spectra are gapped. The 
vacuum values of the ordered fields are $\langle \phi_{s} \rangle =0$ and 
$\langle \phi_{c} \rangle =\sqrt{\pi/8K_{c}}$ and the system has 
long-range CDW order.
Thus the set of vacua (\ref{vacDelta1a}) corresponds to the BI phase. 

For $U^{c}_{ch}<U<U^{c}_{sp}$ the vacuum 
expectation values of the ordered charge field is now different: 
$\langle \phi_{s} \rangle =0$ and $\langle \sqrt{2\pi K_{c}}\phi_{c} \rangle =
\varphi_{0}$. In this phase, long-range CDW and BOW correlations
coexist,
\bea
\nonumber
\langle \Delta_{\small CDW}(x) \Delta_{\small CDW}(x') \rangle 
&\sim&  \sin^{2}\varphi_{0}\, ,\\
\nonumber
\langle \Delta_{\small BOW}(x) \Delta_{\small BOW}(x') \rangle 
&\sim&  \cos^{2}\varphi_{0}\,\, .
\eea  

The standard BOW phase violates site- but preserves link-inversion symmetry. 
In the CDW + BOW phase, due to the CDW pinning by the ionic distortion, the 
link-parity is also broken. However, in the absence of an externally imposed 
Peierls distortion, the BOW pattern is doubly degenerate. Thus the charge 
distribution in the CDW + BOW phase can be represented as a linear 
combination of two dimerized patterns shifted with respect to each other by 
one lattice spacing (see Fig. \ref{fig:SDI}) for which the site-inversion 
symmetry is not broken. 

Finally, at $U>U^{c}_{sp}$ the spin gap disappears. According to FGN at this 
point $\varphi_{0}=0$ and the standard MI phase, with an identical power-law
decay of the SDW and Peierls correlations is realized. 

However, the results of our numerical studies presented below as well as 
QMC calculations \cite{Wilkens00} and finite cluster studies 
using valence bond techniques \cite{Pati00} indicate a different scenario.  
The amplitude of the CDW correlations smoothly decays with 
increasing $U$ with $\varphi_0\rightarrow 0$ for $U\rightarrow\infty$.

\begin{figure}
\centerline{\psfig{file=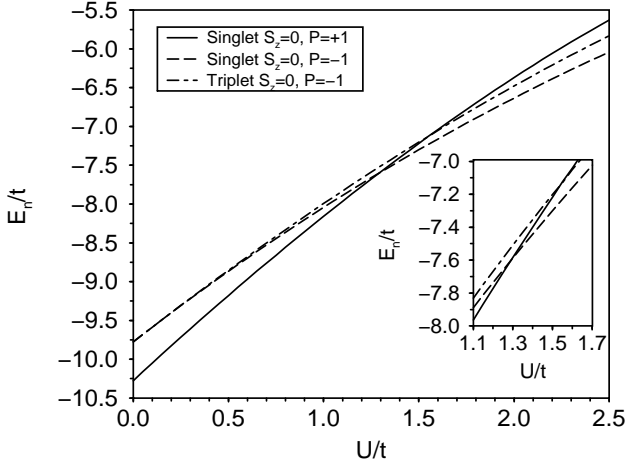,width=85mm,silent=}}
\vspace{2mm}
\caption{Lowest energy eigenvalues of the IHM at 
half-filling for $L=8$ sites, periodic boundary conditions and $\Delta=0.5t$.}
\label{fig:Energies}
\end{figure}

\section{Lanczos exact diagonalization results}

In order to explore the nature of the spectrum and the phase transition, 
we have diagonalized numerically small systems by the Lanczos method
\cite{Lanczos50} extending earlier exact diagonalization calculations 
\cite{Resta95,Gidopoulos99}. The energies of the few lowest eigenstates were 
obtained for finite chains with $L=4n$ 
and PBC or $L=4n+2$ with APBC, for reasons discussed above. 

We first analyze a short chain without finite-size scaling. For chain lengths 
$L\le 16$ finite-size effects 
do not change the qualitative behavior discussed below. In Fig. 
\ref{fig:Energies}, the lowest eigenenergies of the IHM for $\Delta=0.5t$, 
$L=8$ and PBC are shown as a function of $U$. At $U=1.3t$, a level crossing of 
the two lowest eigenstates occurs. A non-degenerate eigenstate of the IHM has
site-parity eigenvalues $\pm 1$, so a ground state level-crossing transition 
corresponds to a change of the site-parity eigenvalue. 

For $U=0$, the IHM is easily diagonalized in momentum space by introducing 
fermionic creation (annihilation) operators 
$\gamma^{\dagger}_{k\sigma b}$ ($\gamma^{\phantom{\dagger}}_{k\sigma b}$)
with an index $b=1,2$ denoting the lower and upper bands, respectively, giving 
two energy bands $E_{1/2}(k)=\pm\sqrt{4\cos^2(k)+(\Delta/4)^2}$ with momenta 
$-\pi/2<k\le\pi/2$. For $U=0$ the first two degenerate 
excited states at half-filling always have negative site parity, because the 
ground state has $P=+1$, and the 
operator $\gamma^{\dagger}_{q\sigma 2}\gamma^{\phantom{\dagger}}
_{q\sigma 1}$ with $q=\pi/2$ obeys 
\begin{equation}
P\gamma^{\dagger}_{q\sigma 2}
 \gamma^{\phantom{\dagger}}_{q\sigma 1}
=-\gamma^{\dagger}_{q\sigma 2}
  \gamma^{\phantom{\dagger}}_{q\sigma 1}P
 \quad.
\end{equation}

\begin{figure}[t!]
\centerline{\psfig{file=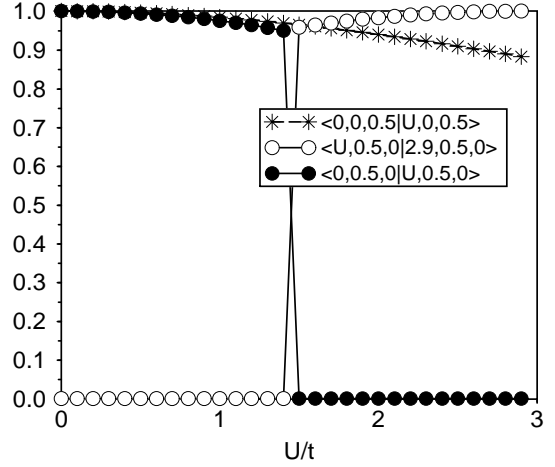,width=85mm,silent=}}
\vspace{2mm}
\caption{Overlap matrix elements of the exact ground states 
$|\psi_0\rangle=|U/t,\Delta/t,\delta\rangle$ of the Peierls chain 
($U=\Delta=0$) and the Peierls-Hubbard model (stars), the ionic band insulator 
($\Delta=0.5t$, $\delta=0$) and the IHM (full circles) and of the IHM at a 
given $U$ and at $U=2.9t$ (circles) as a function of $U$. Calculations were 
performed for $10$ sites with antiperiodic boundary conditions.}
\label{fig:Overlaps}
\end{figure}

The first two excited states shown in Fig. \ref{fig:Energies} are the spin  
singlet ($S=0$, $S_z=0$) and triplet excitations ($S=1$, $S_z=0$), created 
from the ground state by applying the operators 
\begin{eqnarray}
 \nonumber
 &&\frac{1}{\sqrt{2}}\left(
 \gamma^{\dagger}_{q\uparrow 2}
 \gamma^{\phantom{\dagger}}_{q\uparrow 1}
 -\gamma^{\dagger}_{q\downarrow 2}
  \gamma^{\phantom{\dagger}}_{q\downarrow 1}\right)\,\, ,\\
 &&\frac{1}{\sqrt{2}}\left(
 \gamma^{\dagger}_{q\uparrow 2}
 \gamma^{\phantom{\dagger}}_{q\uparrow 1}
 +\gamma^{\dagger}_{q\downarrow 2}
  \gamma^{\phantom{\dagger}}_{q\downarrow 1}\right)
 \quad,
\end{eqnarray}
respectively. Thus both excited states have total momentum $k_{tot}=0$ and 
negative site parity. For $U>0$, these degenerate excited states split in
energy.

Fig. \ref{fig:Overlaps} shows the overlap matrix elements between the exact 
ground states $|\psi_0\rangle=|U/t,\Delta/t,\delta\rangle$ of the IHM, the 
PHM, and the Hubbard model. Specifically we show the overlap 
integrals between the ground states of the IHM for $0<U<2.9t$ and $U=0$, and 
between the ground states for $0<U<2.9t$ and $U=2.9t$. Clearly, a sharp 
transition occurs at $U_c=1.5t$ with the ground states $|U,0.5,0\rangle$ for 
$U<U_c$ and $U>U_c$ being orthogonal to each other. This 
reflects the level-crossing phenomenon with a site-parity change. For the 
PHM ($\Delta=0$, $\delta=0.5$) the situation is different. 
The overlap integral between the ground states for $U=0$ and $U>0$ decreases 
continuously with increasing $U$, and no 
indication of a phase transition is observed. This model is a BI only at $U=0$,
and turns into a correlated Peierls insulator for any $U>0$.


Obviously, exact diagonalization of finite rings identifies one critical
$U_c>0$, separating a BI with $P=+1$ at $U<U_c$ from a CI with $P=-1$ for 
$U>U_c$.

\section{DMRG results}

In order to access the transition scenario in the long chain-length limit, we 
have studied chains up to L=300 using the DMRG method \cite{White92,Peschel99}.
The fact that the transition at $U_c$ is connected with inversion symmetry
requires some caution when open boundary conditions (OBC) are used in our DMRG 
studies. For OBC and $L=2n$ the IHM is not reflection symmetric at any site. 
Thus, the ground state does not have a well defined site-parity, and the 
level-crossing transition is absent.
To overcome this problem, one might try to use chains with OBC and 
an {\it odd} number of sites $L=2n+1$, since the Hamiltonian in this case is 
reflection-symmetric with respect to the site $i_c$ in the center of the chain,
and a site-parity operator is well defined by 
\begin{equation}
 Pc^{\dagger}_{i_c\sigma}P^{\dagger}=c^{\dagger}_{L+1-i_c\sigma}\,\, .
\end{equation}
To test whether this is an improved choice we have calculated the site-parity
of the ground state for $U=0$ analytically for different chain lengths 
$L=2n+1$ and found 
\begin{equation}
 P|\psi_0\rangle=(-1)^n|\psi_0\rangle
 \quad.
\end{equation}
On the other hand, in the large $U$ limit again the mapping to the Heisenberg 
Hamiltonian could be used to obtain the site-parity eigenvalue. By extending 
the idea of Gidopoulos {\it et al.} \cite{Gidopoulos99} to chains with 
$L=2n+1$, for $U\gg t$ we obtain
\begin{equation}
 P|\psi_0\rangle = (-1)^{\left[\sum^{L-1}_{m=1}m\right]}|\psi_0\rangle
                 = (-1)^{n}|\psi_0\rangle
 \quad.
\end{equation}
Thus, the parity eigenvalue of the ground state is the same at $U=0$ and 
$U\gg t$ for a given chain length, and no level crossing occurs. As mentioned 
above, the argument for $U\gg t$ requires some caution for the IHM, so we also 
have checked this result numerically by exact diagonalization of small systems 
with $L=5,7,9,\dots$ to confirm the absence of a level crossing.
Obviously, results for any choice of boundary conditions should recover the 
level-crossing scenario when extrapolated to the thermodynamic limit. Due to 
the fact that the sharp transition at a well defined $U_c$ does not exist in 
the finite-chain results for OBC, the extrapolation is a rather subtle 
problem, since a sharp transition feature has to be identified from the 
extrapolation of smooth curves. This requires the use of quite long chains in
the critical region.

\begin{figure}[t!]
\centerline{\psfig{file=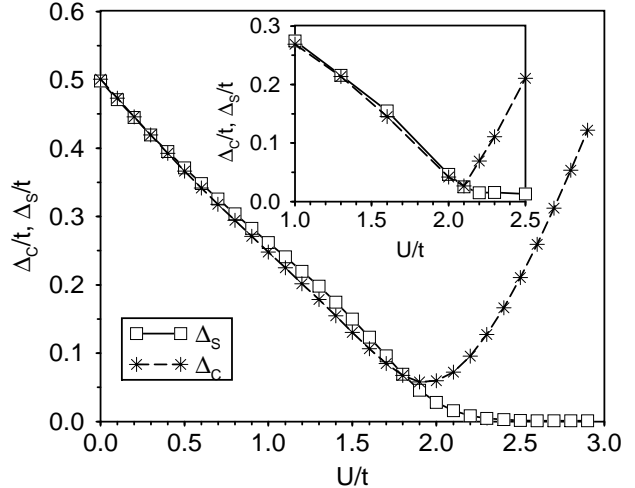,width=90mm,silent=}}
\vspace{2mm}
\caption{Results for the spin ($\Delta_S$) and charge ($\Delta_C$) gaps of the 
IHM at half-filling with $\Delta=0.5t$ as a function of $U$. 
Energies were obtained by DMRG calculations on open chains with 
$L=\{30,40,50,60\}$ (main plot) and $L=\{30,40,50,60,200,300\}$ (inset), and 
extrapolated to the limit of infinite chain length.}
\label{fig:UMuGaps}
\end{figure}

We have evaluated $\Delta_C$ and 
$\Delta_S$ for chains up to 300 sites. In Fig. \ref{fig:UMuGaps} extrapolated 
results are shown as a function of $U$. Calculations were performed with 
OBC for chains of lengths $L=\{30,40,50,60\}$, and additionally for $L=200$ 
and $L=300$ in the transition region around the estimated $U_c$. 
Unlike the definition (\ref{Gaps1}), $\Delta_C$ was obtained here using 
\begin{eqnarray}
 \nonumber
 \Delta_C &=& \frac{1}{2} \Bigl[ E_0(N=L+2,S_z=0)\\\nonumber
          &+& E_0(N=L-2,S_z=0)-2E_0(N=L,S_z=0)\Bigr].
 \label{Gaps2}
\end{eqnarray}
in order to calculate the relevant energies $E_0$ in the subspace
$S_z=0$. This becomes equivalent to (\ref{Gaps1}) in the thermodynamic limit.
We assume a scaling behavior of $\Delta_C$ and $\Delta_S$ of the form 
\cite{Noack00}
\begin{equation}
\Delta_i(L)=\Delta^{\infty}_i+\frac{A_i}{L}+\frac{B_i}{L^2}+\dots
\quad,
\label{GapScaling}
\end{equation}
where $i\in\{S,C\}$. The extrapolation for $L\rightarrow\infty$ is then 
performed by fitting this polynomial in 
$1/L$ to the calculated finite-chain results. We note that different 
finite-size scaling formulas were proposed in the literature mainly when PBC or
APBC were used \cite{Gidopoulos99}.

\begin{figure}[t!]
\centerline{\psfig{file=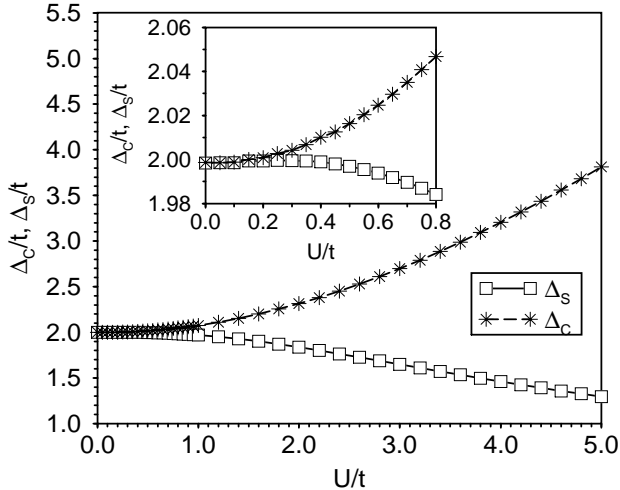,width=90mm,silent=}}
\vspace{2mm}
\caption{Results for the spin ($\Delta_S$) and charge ($\Delta_C$) gaps of the 
Peierls-Hubbard model at half-filling with a modulation of the hopping 
amplitude $\delta=0.5$ as a function of $U$. Energies were obtained by DMRG 
calculations on open chains with 
$L=\{30,40,50,60\}$.}
\label{fig:GapsPeirHub}
\end{figure}

As can be seen from the main plot in Fig. \ref{fig:UMuGaps}, extrapolating the 
results for $L=\{30,40,50,60\}$ does indeed not give a sharp transition 
behavior of the gaps. As illustrated in the inset, 
adding results for $L=200$ and 300 in the critical region changes the picture 
considerably. Within numerical accuracy 
the charge and spin gaps remain equal up to a critical $U_c\approx 2.1t$. A
sharp kink for $\Delta_C$ is observed reflecting the merging of the ground
states of the different site-parity sectors. Importantly, $\Delta_C$ does 
not close at the critical point. This is in fact not in conflict with an 
underlying ground-state level crossing. If the ground states of the different 
site-parity sectors become degenerate, the only rigorous consequence is the 
closing of the optical excitation gap. The selection rules for optical 
excitations (see section VII) allow only for transitions between states 
of different site-parity. Furthermore, optical 
transitions occur within the same particle number sector. The optical gap is 
therefore by definition distinct from the charge gap Eq. (\ref{Gaps1}) which 
involves the removal or the addition of a particle. The critical point $U_c$ 
of the IHM corresponds to the remarkable situation where the optical gap 
closes while $\Delta_C$ remains finite. This implies that each site-parity 
sector separately has a finite $\Delta_C$. Above $U_c$ the charge and spin 
gaps split indicating that the corresponding 
insulating phase is no longer a BI. $\Delta_S$ continuously decreases with 
increasing $U$ and becomes unresolvably small within the achievable numerical 
accuracy.

Our results are in general agreement with the data obtained by Qin {\it et al.}
\cite{Qin00}. These authors performed DMRG calculations for the IHM with 
$\Delta=0.6t$, for chains up to $L=600$ sites. In contrast to our calculations,
Qin {\it et al.} used the definition (\ref{Gaps1}) to calculate $\Delta_C$. 
Surprisingly, they observed a non-monotonic scaling behavior of $\Delta_S$ with
$L$ for values of $U$ close to the critical $U_c$, i.e. for chain lengths 
$L>300$ $\Delta_S$ started to increase again. It remains unclear whether this 
is due to loss of DMRG accuracy with increasing chain lengths. We note that in 
their substantially revised paper Qin {\it et al.} have presented additional 
DMRG data which agree with our present finding. DMRG calculations for the IHM 
with $\Delta=t$ have also been performed by Takada and Kido for chains up to 
$L=400$ sites \cite{Takada00}. The authors interpret their results in the 
region close to $U_c$ in favour of a 
two-transition scenario similar to that of FGN \cite{Fabrizio99}. 
Below we will show that such an interpretation is not valid.

\begin{figure}[t!]
\vspace{5mm}
\centerline{\psfig{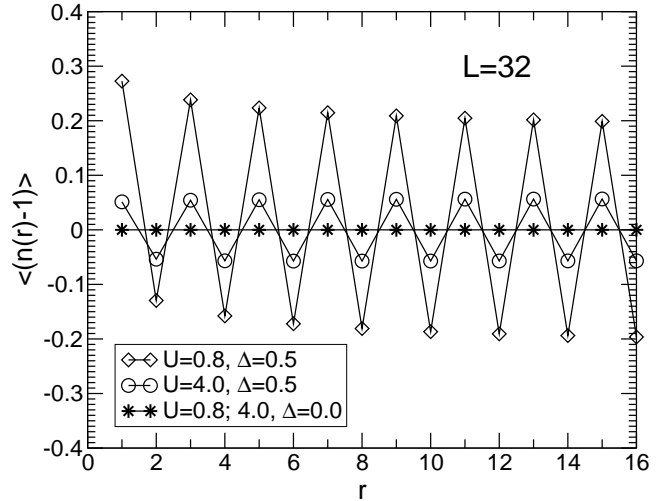}}
\vspace{4mm}
\caption{The electron density distribution in the ground state of the IHM for
$\Delta=0.5t$ and $U=0.8t$ (diamonds) and $U=4t$ (circles). Results were
obtained by  DMRG calculations on open chains with $L=32$. Stars show results
for the Hubbard chain ($\Delta=0$). }
\label{fig:NewCDW}
\end{figure}

For comparison we show in Fig. \ref{fig:GapsPeirHub} the spin and charge gaps
versus $U$ in the PHM. As already
anticipated in previous sections this model does not show any signature of a
phase transition; i.e. $\Delta_c>\Delta_s>0$ for all
$U>0$. So although the Peierls and the ionic insulator for $U=0$ similarly
possess an excitation gap at the BZ boundary, applying a Coulomb $U$ leads to
distinctly different behavior in both cases. The origin of the different
behavior must be traced to the fact that the Hubbard interaction and the ionic
potential compete locally on each site, while the Peierls modulation of the
hopping amplitude tends to move electronic charge to the bonds between sites,
thereby avoiding conflict with
the Hubbard term. This difference is also reflected in the characteristic
structures of the ionic and the Peierls terms in the bosonized versions of the
Hamiltonian.

The important question remains about the nature of the insulating phase of the
IHM for $U>U_c$. As we argued above the numerical search for the vanishing of
the spin gap will remain ambiguous due to the necessity to rule out the
possibility of a finite but exponentially small $\Delta_S$. To further analyze
the BI and CI phases below and above $U_c$, we have
calculated correlation functions in the ground state of the IHM using DMRG
results for finite chains.

\begin{figure}[t!]
\centerline{\psfig{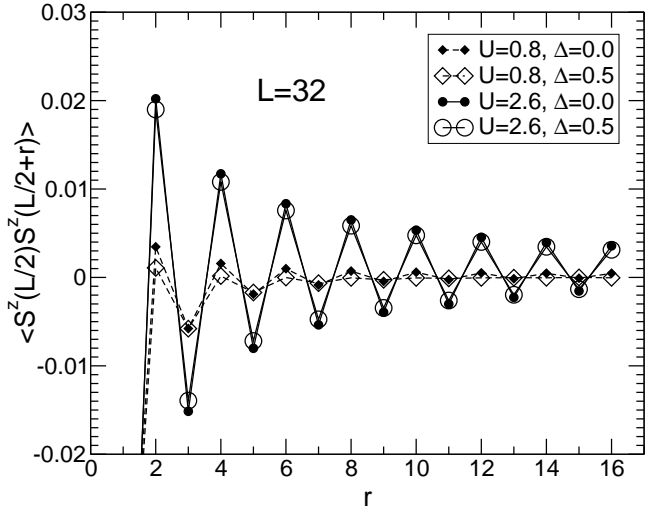}}
\vspace{4mm}
\caption{Spin-spin correlation function in the ground state of the IHM for
$\Delta=0.5t$ (open symbols) and the Hubbard model ($\Delta=0$) (full symbols)
at $U=0.8t$ (diamonds) and $U=4t$ (circles). Chain lenght $L=32$.}
\label{fig:NewSDW}
\end{figure}

Fig. \ref{fig:NewCDW} shows the charge distribution $\langle 0|
(n_{r,\uparrow} + n_{r,\downarrow} -1)|0 \rangle $ in the ground state of the
Hubbard model and the IHM at $U=0.8t<U_c$ and $U=4t>U_c$ for a $L=32$ chain.
The alternating pattern in the density distribution is well pronounced not only
in the BI phase but also in the CI phase far beyond  the critical point
at $U\gg U_{c}$. For the $L=32$ chain the CDW is well established at distances 
$l \sim L/2$ even at $U=4t$. We note, that the attraction (or repulsion)
of the charge from the chain edges, a well
pronounced boundary effect in the BI phase, is absent at $\Delta=0$ and in the
CI phase of the IHM. The amplitude of the CDW pattern smoothly decreases with
increasing $U$. Our numerical data indicate that the alternating pattern
in the electron density distribution in the IHM remains for arbitrary
finite $U$.  

Fig. \ref{fig:NewSDW} shows the DMRG results for the spin-spin correlation 
function $\langle 0|S^{z}(L/2)S^{z}(L/2+r)|0\rangle$ in the 
ground state of the IHM at $U=0.8t<U_c$ and $U=4t>U_c$ in comparison with the
spin correlators in the Hubbard model. In the BI phase at $U=0.8t$ 
the SDW correlations are almost completely suppressed on a scale of half the 
chain length. At $U=4t$ the amplitude of the SDW correlation in the CI phase 
of the IHM is slightly reduced in comparison to the Hubbard model at the 
same value of $U$. However, the large distance behavior of the
spin correlations in the CI phase and the MI phase of the Hubbard model is
similar. Therefore at arbitrary $U>U_c$ the spin correlations in the CI phase
of the IHM are almost identical to the Hubbard model. This equivalence,
however, is valid up to the accuracy of an unresolvably small gap in the
spin excitation spectrum.

\begin{figure}[t!]
\centerline{\psfig{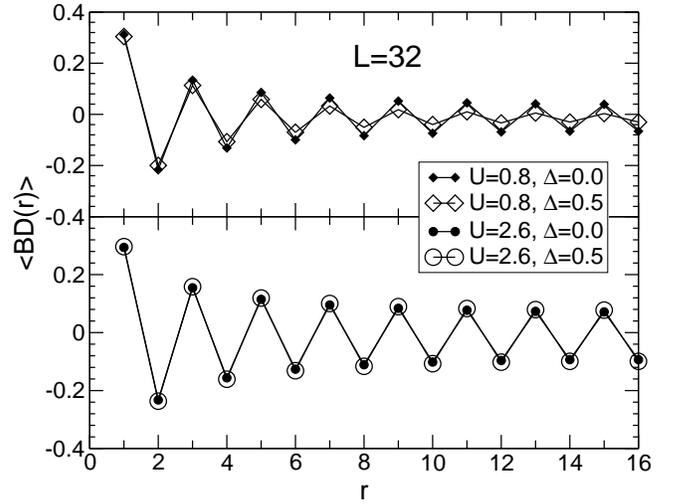}}
\vspace{4mm}
\caption{Bond density correlation function in the ground state of the IHM for
$\Delta=0.5t$ (open symbols) and the Hubbard model (full symbols) at $U=0.8t$
(diamonds) and $U=4t$ (circles).}
\label{fig:NewBDW}
\end{figure}

To address the dimerization tendencies in the CI phase we have calculated the
ground-state distribution of the normalized bond density
\begin{equation}
BD(r) = \frac{\sum_{\sigma}\langle c^{\dagger}_{r,\sigma}c^{\phantom{\dagger}}_{r+1,\sigma} + H.c.\rangle}{\frac{1}{L}\sum_{r,\sigma}\langle c^{\dagger}_{r,\sigma}c^{\phantom{\dagger}}_{r+1,\sigma} + H.c.\rangle}-1\,.
\end{equation}
Fig. \ref{fig:NewBDW} shows the results of the DMRG calculations for a $L=32$
IHM chain and the Hubbard chain at $U=0.8t$ and $U=4t$. The boundary effect of
an open chain is strong and leads to a modulation of the bond density already
for the pure Hubbard model. The comparison with the Hubbard chain indicates, 
that the ionic distortion leads to a weak suppression of the bond-density 
ordering at $U<U_{c}$, while at $U > U_{c}$, the amplitude of the bond-density 
modulations slightly increases in the CI phase.

\begin{figure}[t!]
\centerline{\psfig{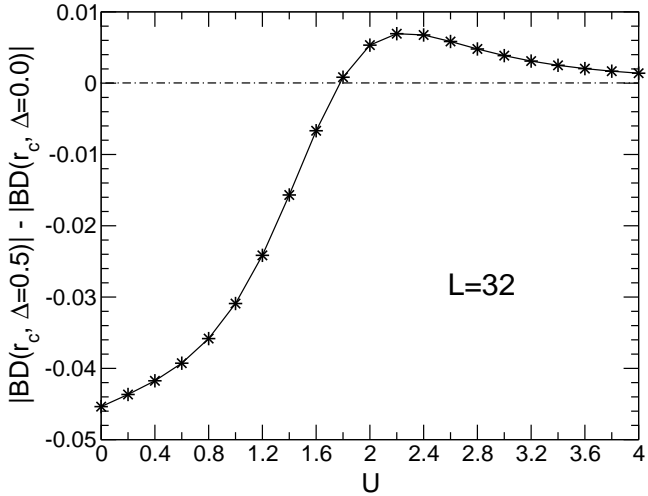}}
\vspace{4mm}
\caption{Difference between the bond-charge density of the IHM with
$\Delta=0.5t$ and the Hubbard model for $\Delta=0$ as a function $U$. $r_c$
denotes the bond $\langle r_c,r_c+1\rangle$ in the center of a $L=32$ chain.}
\label{fig:OBD}
\end{figure}

To study the tendency in the IHM towards BOW ordering at the
transition into the CI phase we compare the boundary induced alternating
patterns of the bond density at $\Delta=0$ and $\Delta=0.5t$ as a function of
$U$. For the open $L=32$ IHM chain the effect is clearly seen if we compare the
corresponding bond densities in the central bond of the chain as shown in Fig.
\ref{fig:OBD}. In the BI phase at $U<U_{c}$, the ionic term reduces the bond
density. However in the vicinity of the transition point and at $U>U_c$ the
bond density in the IHM is larger than in the pure Hubbard model. It is notable
that this difference remains positive for arbitrary $U>U_c$ and disappears only
in the limit $U\rightarrow \infty$.

We conclude that the finite chain DMRG studies of the
IHM indicate the presence of only one transition from the BI to the
unconventional CI phase with long range CDW order for all $U<\infty$. In the CI
phase the spin excitation spectrum is either gapless or characterized by an 
unresolvably small spin gap. If the spin gap is finite, we expect true long 
range BOW order coexisting with a CDW. If however the spin gap vanishes, 
SDW and the dimer-dimer correlations coexist in this phase with an identical 
decay at large distances in the presence of a finite CDW amplitude. The fact, 
however, that the spin gap is finite and equal to $\Delta_C$ at the transition 
supports the scenario of a CI phase for $U>U_c$ with a finite $\Delta_S$ and 
coexisting long range CDW and BOW order.

\section{Optical Conductivity}

To further analyze the BI and CI phases we have 
calculated the frequency-dependent conductivity within the dynamical 
DMRG (DDMRG) approach, making use of the correction-vector technique 
described in Ref. \cite{Kuehner99}. Calculations were performed for $L=128$ 
sites with OBC using a Parzen filter to suppress the influence of the chain 
boundaries.

The real part of the dynamical conductivity $\sigma(\omega)$ is determined 
from the linear response of the system to an external electromagnetic field. In
the Kubo formalism, $\sigma(\omega)$ is related to the imaginary part of the 
retarded current-current correlation function
\begin{eqnarray}
 \nonumber
 \chi_{jj}(q,\omega)
 &=&{i \over La}\int^{\infty}_{0}d{\tau}\,e^{i\omega \tau}\,
                \langle\psi_0|\,[j_{-q}(\tau),j_q(0)]\,|\psi_0\rangle\\
 \nonumber
 &=&{\hbar\over La}\sum_{n\neq 0}\left(
    {|\langle\psi_n|j_q|\psi_0\rangle|^2\over\hbar\omega+(E_n-E_0)+i0^+}
    \right.\\
  &&\phantom{{\hbar\over La}\sum_{n\neq 0}}\left.
    -{|\langle\psi_n|j_q|\psi_0\rangle|^2\over \hbar\omega-(E_n-E_0)+i0^+}
    \right)
 \quad,
 \label{currentcorr}
\end{eqnarray} 
with $\tau$ the time index, $a$ the distance between neighboring sites, and 
$|\psi_n\rangle$ and $E_n$ denoting the eigenstates and their respective 
energies. The paramagnetic current operator is defined by
\begin{equation}
 j_q = -it{ea\over\hbar}\sum_{l,\sigma} e^{iql}
                                    \left(c^{\dagger}_{l+1\sigma}
                                          c^{\phantom{\dagger}}_{l\sigma}
                                         -c^{\dagger}_{l\sigma}
                                          c^{\phantom{\dagger}}_{l+1\sigma}
                                    \right)
 \quad,
\end{equation} 
where $e$ is the electron charge. The structure of the current operator and 
the required matrix elements enforces the important selection rule that only 
transitions between states with different site-parity are allowed. The real 
part of the conductivity in the long-wavelength limit $q=0$ is 
given by
\begin{eqnarray}
 \sigma_1(\omega)&=&D\delta(\omega)+\sigma^{reg}_1(\omega>0)\\
 \sigma^{reg}_1(\omega>0)&=&{1\over\hbar\omega}\,{\mathrm{Im}}
                          \chi_{jj}(q=0,\omega)
 \quad.
\end{eqnarray} 
For the insulating phases of the IHM at half-filling the Drude weight either 
vanishes, $D=0$, or is not defined for degenerate ground states, so that 
$\sigma_1(\omega)=\sigma^{reg}_1(\omega)$. 

\begin{figure}[t]
\centerline{\psfig{file=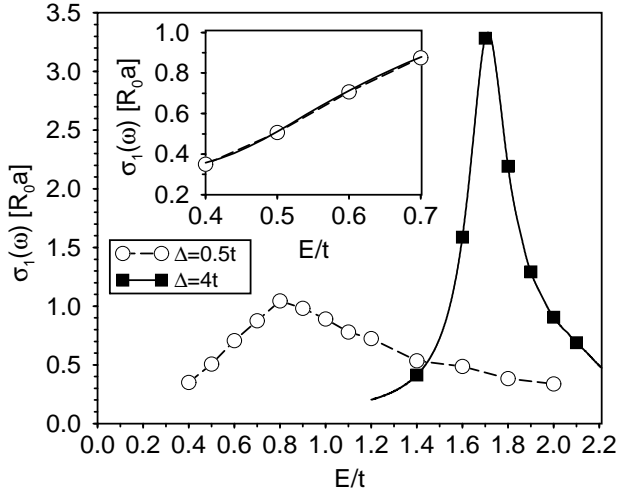,width=90mm,silent=}}
\vspace{2mm}
\caption{Real part of the optical conductivity $\sigma_1(\omega)$ in units of
$R_0a=e^2a/\hbar$ vs. energy $E=\hbar\omega$ for the half-filled IHM for 
$U=3t$. The results were obtained by DDMRG calculations using the 
correction-vector method for a chain with $L=128$ sites and OBC. A finite 
broadening $\eta=0.1t$ has been used. The selected $\Delta$-values correspond 
to the BI ($\Delta=4t$) and the CI ($\Delta=0.5t$) phase. The inset shows
a fit (solid black line) to the onset region of $\sigma_1(\omega)$ in the CI 
phase.}
\label{fig:OptCond1}
\end{figure}

In the absence of interactions $U=0$, the conductivity diverges as
$\sigma_1(\omega)\sim 1/\sqrt{\omega-\Delta_C}$ for $\omega\rightarrow\Delta_C$
and $\omega>\Delta_C$ \cite{Gebhard97}. We expect this behavior to persist upon
increasing $U$ over the entire BI phase. In order to test this expectation we 
have evaluated the optical conductivity in the BI phase by DDMRG 
\cite{Kuehner99}. For the correction-vector method a separate DMRG run has to 
be performed for each selected frequency $\omega$. In Fig. \ref{fig:OptCond1}, 
the DDMRG results for a chain with $L=128$ sites and OBC are plotted for 
$U=3t$. For the proper convergence of the correction-vector algorithm a finite 
broadening $\eta=0.1t$ was chosen in Eq. (\ref{currentcorr}). For $\Delta=4t$,
the solid black line interpolating between the discrete data points at 
different frequencies was calculated by additionally applying the Lanczos 
vector method after the DMRG sweeps, obtaining the correlation function also in
the vicinity of the selected frequencies \cite{Kuehner99}. One observes a 
dominant low-energy 
excitation peak above the optical gap at around $E\approx 1.7t$. The Lorentzian
tail for $E<1.7t$ results from the necessarily large broadening. Also, in the 
correction-vector results the square-root divergence at the absorption edge is 
not visible due to the finite size of the system. However, it is to be 
anticipated from the dominant excitation peak above the gap edge. 

For $U>U_c$, the situation is different, as demonstrated in Fig. 
\ref{fig:OptCond1}, too, where we also show $\sigma_1(\omega)$ for $U=3t$ and 
$\Delta=0.5t$. The dashed line connecting the data points serves only as a 
guide to the eye. In the CI phase, the behavior of the optical excitations is 
expected to be similar to those of the repulsive Hubbard model, where 
$\sigma_1(\omega)\sim{\sqrt{\omega-\Delta_C}}$ above the charge gap 
$\Delta_C$ which in this case is identical to the optical gap 
\cite{Jeckelmann99}. In the inset of Fig. \ref{fig:OptCond1}, 
$\sigma_1(\omega)$ is plotted on a finer energy scale in the onset region above
the gap edge. The solid black line is a fit to the DDMRG data obtained by 
convoluting $\sqrt{\hbar\omega-\Delta_{opt}}$ with a Lorentzian of width 
$\eta$, i.e. by using the fit formula
\begin{equation}
 \sigma_1(\omega)={A \over \hbar\omega\pi}\int^{\nu_{max}}_{\Delta_{opt}}
     {\rm d}\nu\,
                  {\eta\sqrt{\nu-\Delta_{opt}}\over(\nu-\hbar\omega)^2+\eta^2}
 \quad,
\end{equation}
where $A$ is an adjustable prefactor and $\nu_{max}$ the upper bound for the 
square-root dependence. The fit in Fig. \ref{fig:OptCond1} has been obtained 
using the parameter values $A=1.9t$, $\nu_{max}=1.5t$, and 
$\Delta_{opt}=0.48t$. The value for $\Delta_{opt}$ is close to the charge gap 
result presented in Fig. \ref{fig:UMuGaps}. A difference between 
$\Delta_{opt}$ and $\Delta_C$ can not be resolved within the accuracy of the 
DDMRG data. The agreement between the fit and the DDMRG results is quite 
striking. Thus, we conclude that the numerical results are in 
agreement with the behavior expected for a correlated (Mott) insulator in 1D. 
The qualitative behavior of $\sigma_1(\omega)$ compares well with the results 
obtained for the Hubbard model at $U=3t$ \cite{Jeckelmann99}. 

Further consequences for the optical conductivity follow from the level 
crossing scenario discussed in the previous sections. On approaching $U_c$ the 
ground and the first excited state become degenerate, $E_1\rightarrow E_0$; the
matrix element $\langle\psi_1|j_{q=0}|\psi_0\rangle$ between these states
remains finite due to their different site parity eigenvalues. Therefore, the 
optical gap vanishes precisely at $U_c$. The critical point has been commonly 
referred to in the literature 
as ``metallic''. However, by definition a metal is characterized not only 
by gapless excitations, but also by a finite Drude weight $D>0$. Due to
the ground-state degeneracy the Drude weight for $U=U_c$ is not properly 
defined. Thus it follows that although the optical excitation gap vanishes in 
the IHM for $U=U_c$, it cannot be considered ``metallic'' in the 
usual sense. (Following the definition of W. Kohn a metal has a finite Drude 
weight, $D>0$ \cite{Kohn}.)

We emphasize that the optical conductivity was so far evaluated in the BI and 
the CI phase sufficiently far from the critical point. Our currently achievable
accuracy does not allow safe conclusions for the optical conductivity in the 
critical region around $U_c$. At $U_c$ the charge gap $\Delta_C$ remains finite
while the optical gap vanishes. This special situation demands that near below 
$U_c$ the optical conductivity has an
isolated peak below the onset of the continuum for optical absorption. The
energy of the isolated peak corresponds to the energy difference of the 
individual ground states in each site-parity sector with $P=\pm 1$. 
Resolving this remarkable feature will remain a challenge for future numerical 
simulations. 

\section{Conclusions}
Our combined analytical and numerical analysis clarifies the ground-state 
structure of the IHM. A single transition at a critical $U_c(\Delta)$ 
separates a CDW-BI from a CI phase with finite charge and optical 
excitation gaps and most likely also a finite spin excitation gap. Above $U_c$,
CDW and BOW order coexist and persist 
for all $U>U_c$. No secondary transition into a true MI phase occurs which - 
in fact - would have a higher symmetry than the IHM itself. The 
insulator-insulator transition on finite chains of the IHM results from a 
ground-state level crossing of the site-parity sectors with $P=\pm 1$. The BI 
phase has a unique ground state with $P=+1$ while the ground state in the 
CDW + BOW phase above $U_c$ is necessarily two-fold degenerate. The critical
point $U_c(\Delta)$ is characterized by the remarkable situation that the 
optical absorption gap vanishes while the spin and charge excitation gaps 
remain finite and equal. The distinction between the optical and the charge gap
is the key feature for the structure of the insulating phases of the IHM.
Refined numerical studies of the optical conductivity in the critical region 
define therefore the future task. 

The existence of an insulator-insulator transition in the IHM at half-filling 
naturally raises the question of whether a ground state phase transition finds 
its continuation also when the system is doped away from half-filling. The 
possible consequences for the dynamics of doped charge carriers and for the 
phenomenon of spin-charge separation will be the topic of future investigations
in the IHM.

\acknowledgements

We thank D. Baeriswyl, M. Fabrizio, A. Nersesyan, B. Normand, H. Fehske, and 
T. Vekua for helpful discussions and R. Noack for instructive comments on the 
DMRG method. This work was supported by the Deutsche Forschungsgemeinschaft 
(DFG) through SP 1073. APK also acknowledges support through 
Sonderforschungsbereich 484 of the DFG. GIJ and MS acknowledge support by the 
SCOPES grant N 7GEPJ62379.

\end{document}